\documentclass[10pt, conference]{IEEEtran}
\IEEEoverridecommandlockouts
\usepackage{graphicx}
\usepackage{svg}
\usepackage{comment}
\usepackage{bbding}
\usepackage{multirow}
\usepackage{longtable}
\usepackage{array}
\usepackage{url} 
\usepackage{cite}
\usepackage{float}
\usepackage{amsmath}
\usepackage{amssymb}
\usepackage{indentfirst}
\usepackage{verbatim}
\usepackage{makecell}
\usepackage{soul}
\usepackage{tabularx}
\usepackage{booktabs}
\usepackage{subcaption}
\usepackage{lettrine}
\usepackage{soul}
\usepackage{algorithm}
\usepackage{algorithmic}
\usepackage{colortbl} 
\usepackage{subcaption}

\usepackage[shortcuts,acronym, nonumberlist]{glossaries}

\def\BibTeX{{\rm B\kern-.05em{\sc i\kern-.025em b}\kern-.08em
    T\kern-.1667em\lower.7ex\hbox{E}\kern-.125emX}}

\makeglossaries 
\newacronym{pi}{PI}{{Patch Independent}}
\newacronym{pd}{PD}{{Patch Dependent}}
\newacronym{cl}{CL}{{Contrastive Learning}}
\newacronym{mlp}{MLP}{{Multi-Layer-Perceptron}}
\newacronym{uwb}{UWB}{{Ultra-Wideband}}
\newacronym{cir}{CIR}{{Channel Impulse Response}}
\newacronym{csi}{CSI}{{Channel State Information}}
\newacronym{cfr}{CFR}{{Channel Frequency Response}}
\newacronym{iq}{IQ}{In-phase and Quadrature}
\newacronym{los}{LOS}{{Line-of-Sight}}
\newacronym{nlos}{NLOS}{{Non-Line-of-Sight}}
\newacronym{lp}{LP}{{Linear Probing}}
\begin{document}

\title{Lightweight Foundation Model for Wireless Time Series Downstream Tasks on Edge Devices \\

\thanks{This work was partly funded by the SNS JU European Union’s Horizon Europe research and innovation program under Grant Agreement No 101139194 (6GXCEL), the Horizon Europe program under the MCSA Staff Exchanges Grant Agreement 101086218 (EVOLVE), the Fund for Scientific Research Flanders (FWO-Vlaanderen) under SB-PhD Fellowship with grant number 1S52025N and FWO research project PESSO with grant number G018522N.}
}

\author{\IEEEauthorblockN{1\textsuperscript{st} Mohammad Cheraghinia}
\IEEEauthorblockA{\textit{IDLab, Ghent University - imec}\\
Ghent 9052, Belgium \\
mohammad.cheraghinia@ugent.be}
\and
\IEEEauthorblockN{2\textsuperscript{nd} Eli De Poorter}
\IEEEauthorblockA{\textit{IDLab, Ghent University - imec}\\
Ghent 9052, Belgium \\
eli.depoorter@ugent.be}
\and
\IEEEauthorblockN{3\textsuperscript{rd} Jaron Fontaine}
\IEEEauthorblockA{\textit{IDLab, Ghent University - imec}\\
Ghent 9052, Belgium \\
jaron.fontaine@ugent.be}
\and
\IEEEauthorblockN{4\textsuperscript{th} Kwang Soon Kim}
\IEEEauthorblockA{\textit{Scho. of Electrical and Electronic Eng.}\\
\textit{Yonsei University}\\
Seoul 03722, South Korea\\
ks.kim@yonsei.ac.kr}
\and
\IEEEauthorblockN{5\textsuperscript{th} Merouane Debbah}
\IEEEauthorblockA{\textit{Center for 6G Technology} \\
\textit{Khalifa University of
Science and Technology}\\
Abu Dhabi, United Arab Emirates \\
merouane.debbah@ku.ac.ae}
\and
\IEEEauthorblockN{6\textsuperscript{th} Adnan Shahid}
\IEEEauthorblockA{\textit{IDLab, Ghent University - imec}\\
Ghent 9052, Belgium \\
adnan.shahid@ugent.be}

}

\maketitle

\begin{abstract}
While machine learning is widely used to optimize wireless networks, training a separate model for each task in communication and localization is becoming increasingly unsustainable due to the significant costs associated with training and deployment. Foundation models offer a more scalable alternative by enabling a single model to be adapted across multiple tasks through fine-tuning with limited samples. However, current foundation models mostly rely on large-scale Transformer architectures, resulting in computationally intensive models unsuitable for deployment on typical edge devices. This paper presents a lightweight foundation model based on simple \ac{mlp} encoders that independently process input patches. Our model supports 4 types of downstream tasks (long-range technology recognition, short-range technology recognition, modulation recognition and line-of-sight-detection) from multiple input types (IQ and CIR) and different sampling rates. We show that, unlike Transformers, which can exhibit performance drops as downstream tasks are added, our \ac{mlp} model maintains robust generalization performance, achieving over 97$\%$ accurate fine-tuning results for previously unseen data classes. These results are achieved despite having only 21K trainable parameters, allowing an inference time of 0.33 ms on common edge devices, making the model suitable for constrained real-time deployments. 
\end{abstract}


\section{Introduction}
The field of wireless communications is currently struggling with an unsustainable paradigm: the necessity of developing, training, and deploying distinct, specialized models for every individual application. This task-specific approach incurs large costs in terms of data acquisition, data labeling, computational resources, and model creation time, creating a major bottleneck for innovation. Foundation models have emerged as a promising, "greener" alternative, offering a unified approach where a single, large-scale model is pre-trained on extensive unlabeled data and subsequently fine-tuned to excel at a wide array of downstream tasks. This approach simplifies deployment and allows the trained foundation model to be shared publicly, broadening access to advanced AI capabilities \cite{Chen_2024_CVPR}.

However, the dominant trend in foundation models has shifted toward large-scale, multi-million-parameter Transformer-based architectures \cite{10.1007/978-3-031-86623-4_12}. While powerful, their immense size and computational demands make them impractical for a critical and rapidly expanding domain: edge computing. Real-world wireless applications, such as real-time spectrum analysis, dynamic network management, and on-device localization, require real-time intelligence directly at the data source \cite{7488250}. Deploying large Transformer models on these resource-constrained edge devices is often infeasible due to limitations in memory, processing power, and energy consumption. This creates a clear and pressing need for a new class of foundation models: ones that are not only powerful and adaptable but also exceptionally lightweight and efficient.

This paper directly addresses this challenge by questioning the "bigger is better" assumption. We introduce a lightweight foundation model that is specifically designed for the demands of edge environments. Our primary contributions are as follows:
\begin{itemize}
    \item We propose a lightweight foundation model built upon a simple \ac{mlp} encoder that utilizes a \ac{pi} pre-training strategy. This approach drastically reduces model complexity compared to conventional \ac{pd} Transformer models, resulting in a model with only $\approx$21K parameters.
    \item We demonstrate through comprehensive experiments that our compact model can be effectively fine-tuned to recognize new, previously unseen wireless technologies. Compared to Transformer models, our lightweight model achieves competitive or superior accuracy across four diverse downstream tasks: short-range technology recognition, long-range technology recognition, line-of-sight detection, and modulation recognition.  
    \item We validate our model's computational efficiency by demonstrating significantly reduced pre-training and inference times compared to its counterparts, highlighting its ability to achieve very low inference times on an NVIDIA Jetson Xavier NX edge device.
\end{itemize}

The remainder of this paper is structured as follows. Section~\ref{sec:related_works} reviews related works in the domain of wireless foundation models. Section~\ref{sec:system_model} describes our system model, datasets, and pre-training strategy. Section~\ref{sec:methodology} provides a detailed breakdown of our proposed methodology and lightweight architecture. Section~\ref{sec:Results} presents the experimental results, including performance benchmarks and complexity analysis. Finally, Section~\ref{sec:Conc} concludes the paper and outlines directions for future work.

\section{Related Works} \label{sec:related_works}
This section discusses recent works that proposed foundation models using wireless data modalities as input, primarily focusing on raw \ac{iq} samples or higher-level representations like \ac{csi} and spectrogram inputs.

SpectrumFM~\cite{zhou2025spectrumfmfoundationmodelintelligent} introduces a large-scale model with approximately 30.7 million parameters, designed to address four downstream tasks: modulation classification, wireless technology classification, spectrum sensing, and anomaly detection. Similarly, IQFM~\cite{mashaal2025iqfmwirelessfoundationalmodel} presents a more lightweight CNN-based model with about 341K parameters for a comparable set of four tasks. While both demonstrate the potential of foundation models for multi-task learning, they are pre-trained on datasets that include all target classes and tasks, without evaluating how well the proposed model generalizes to previously unseen scenarios. In contrast, the work in~\cite{cheraghinia2025foundationmodelwirelesstechnology} proposes a Transformer-based model with $\approx$700K parameters that is evaluated handling unseen classes. These \ac{iq}-based models highlight a trade-off between model size, the number of supported tasks, and the ability to generalize to new classes. 

Beyond using raw \ac{iq} data as input, other works have explored foundation models using different input representations. For example, WiFo~\cite{liu2025wifo} and LWLM~\cite{pan2025largewirelesslocalizationmodel} utilize inputs like \ac{csi} and \ac{cfr}, respectively. These approaches treat channel data as images, applying vision-based techniques such as Vision Transformers (ViT) for tasks like channel prediction and localization. However, this strategy presents two drawbacks for edge deployment. Firstly, it imposes considerable pre-processing. Extracting \ac{csi}/\ac{cfr} from raw radio signals increases both power consumption and latency, directly conflicting with the operational constraints of edge devices, such as limited memory, compute resources, and battery life.  Secondly, it creates a dependency on specific hardware and communication standards that can provide such structured data, limiting the model's generalizability. Our work deliberately avoids this by operating on lower-level data representations (\ac{iq} samples and \ac{cir}), thereby creating a more lightweight, versatile, and efficient model better suited for the demands of edge computing. 

\begin{table}[h]
\centering
\begin{tabular}{|m{1.75 cm}|m{0.75 cm}|m{1.35 cm}|m{1.4 cm}|m{1.3 cm}|}
\hline
\textbf{Paper} & \textbf{Unseen Class} & \textbf{Downstream tasks} & \textbf{Number of parameters} & \textbf{Input} \\
\hline
\hline
WiFo\cite{liu2025wifo} & \checkmark & 2 & 21.6M & CSI \\
\hline

LWLM\cite{pan2025largewirelesslocalizationmodel} & - & 4 &   5.27M & CFR\\

\hline
SpectrumFM\cite{zhou2025spectrumfmfoundationmodelintelligent} & - & 4 & $\approx$30.7M & IQ samples \\
\hline
IQFM\cite{mashaal2025iqfmwirelessfoundationalmodel} & - & 4 & $\approx$341K & IQ samples \\
\hline
Our previous work \cite{cheraghinia2025foundationmodelwirelesstechnology} & \checkmark & 2 & $\approx$700K & IQ samples \\

\hline
\hline
Our work & \checkmark & 4 & $\approx$21K & IQ samples or CIR \\
\hline
\end{tabular}
\caption{Overview of related works on wireless foundation models.}
\label{tab:related_work}
\end{table}

As summarized in Table~\ref{tab:related_work}, our work distinguishes itself by proposing an exceptionally lightweight model with only $\approx$21K parameters, making it particularly well-suited for deployment on resource-constrained edge devices. Despite its small footprint, our model supports four downstream tasks and is designed and evaluated for its ability to generalize to unseen classes. Furthermore, it is unique in that it can process both \ac{iq} samples and \ac{cir} data, making it applicable to a broader range of wireless technologies than prior work.

\begin{figure}[!htbp]
    \centering
    \includegraphics[width=0.95\linewidth]{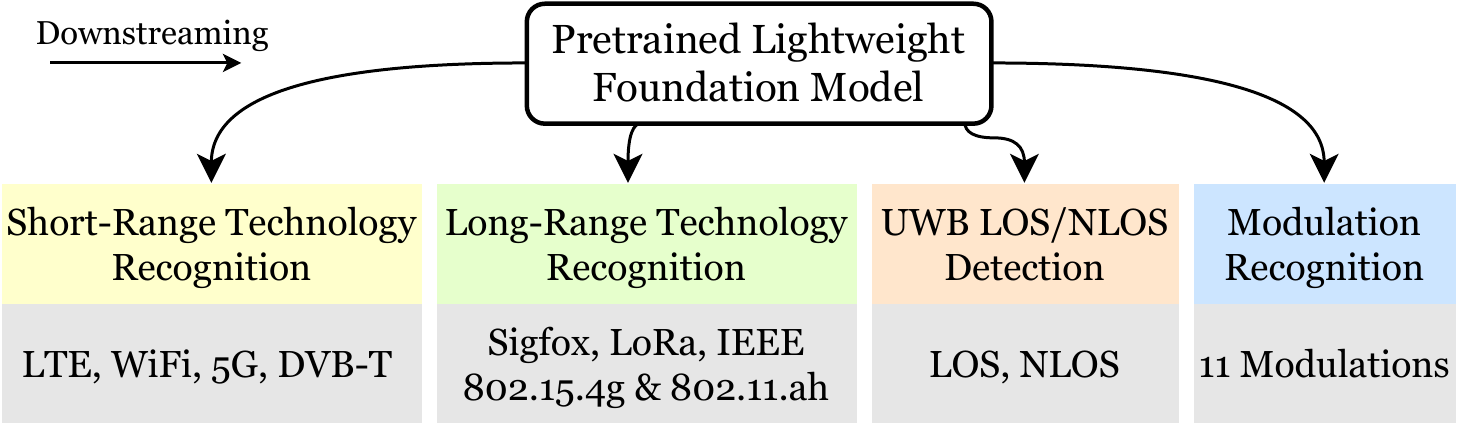}
    \caption{Our pre-trained lightweight foundation model is downstreamed for four distinct downstream tasks in wireless communication.}
    \label{fig:system_model}
\end{figure}

\section{System Model} \label{sec:system_model}
\subsection{System Description}
Our proposed system consists of a foundation model first \textbf{pre-trained} on a comprehensive, unlabeled dataset of long-range (e.g., Sigfox, LoRa) and short-range (e.g., WiFi, 5G) radio signals, \ac{uwb} \ac{los}/\ac{nlos}, and modulation recognition as depicted in Fig~\ref{fig:system_model}. This pre-training utilizes self-supervised learning to build generalizable representations of wireless signals. To evaluate generalization capacity, we consider three scenarios in pre-training: (i) all classes included, (ii) Zigbee excluded, and (iii) LTE excluded.

Following pre-training, the model is \textbf{fine-tuned} using a small, labeled dataset that includes these previously unseen technologies alongside the other classes. This allows the model to extend its recognition capabilities to new wireless standards with minimal additional data, a crucial feature for real-world applications where obtaining large labeled datasets is often impractical and costly.

\subsubsection{Pre-training Phase}

We investigate a self-supervised pre-training strategy to learn initial model parameters ($\theta$).

Using an unlabeled dataset $\mathcal{D}_{\text{pre-training}} = \{X^{(u)}\}_{u=1}^U$, where each sample is indexed by $u$, we employ a patch-level masking strategy for improving generalization, data efficiency, and computational efficiency. A portion of each time-series signal $X^{(u)}$ is masked as an input to the model. The model is then trained to reconstruct the original signal from the masked input. In addition to reconstruction, we apply a contrastive loss to enhance representation learning by bringing similar samples closer together in the embedding space and pushing dissimilar ones apart. This encourages the model to learn meaningful temporal features without relying on labels.

\subsubsection{Fine-tuning Phase}

During fine-tuning, we adapt the pre-trained model to specific downstream classification tasks while ensuring that it can generalize even if some classes were not seen during pre-training. The fine-tuning dataset, $\mathcal{D}_{\text{fine-tuning}}$, is significantly smaller than the pre-training set and includes samples from both the original classes in the pre-training dataset and new, unseen classes like Zigbee or LTE. The datasets are mutually exclusive ($\mathcal{D}_{\text{pre-training}} \cap \mathcal{D}_{\text{fine-tuning}} = \emptyset$) to ensure a fair evaluation of generalization. This approach allows the model to efficiently adapt to new tasks while retaining its foundational understanding of wireless signals, balancing knowledge and flexibility.

\begin{figure}
    \centering
    \includegraphics[width=1\linewidth]{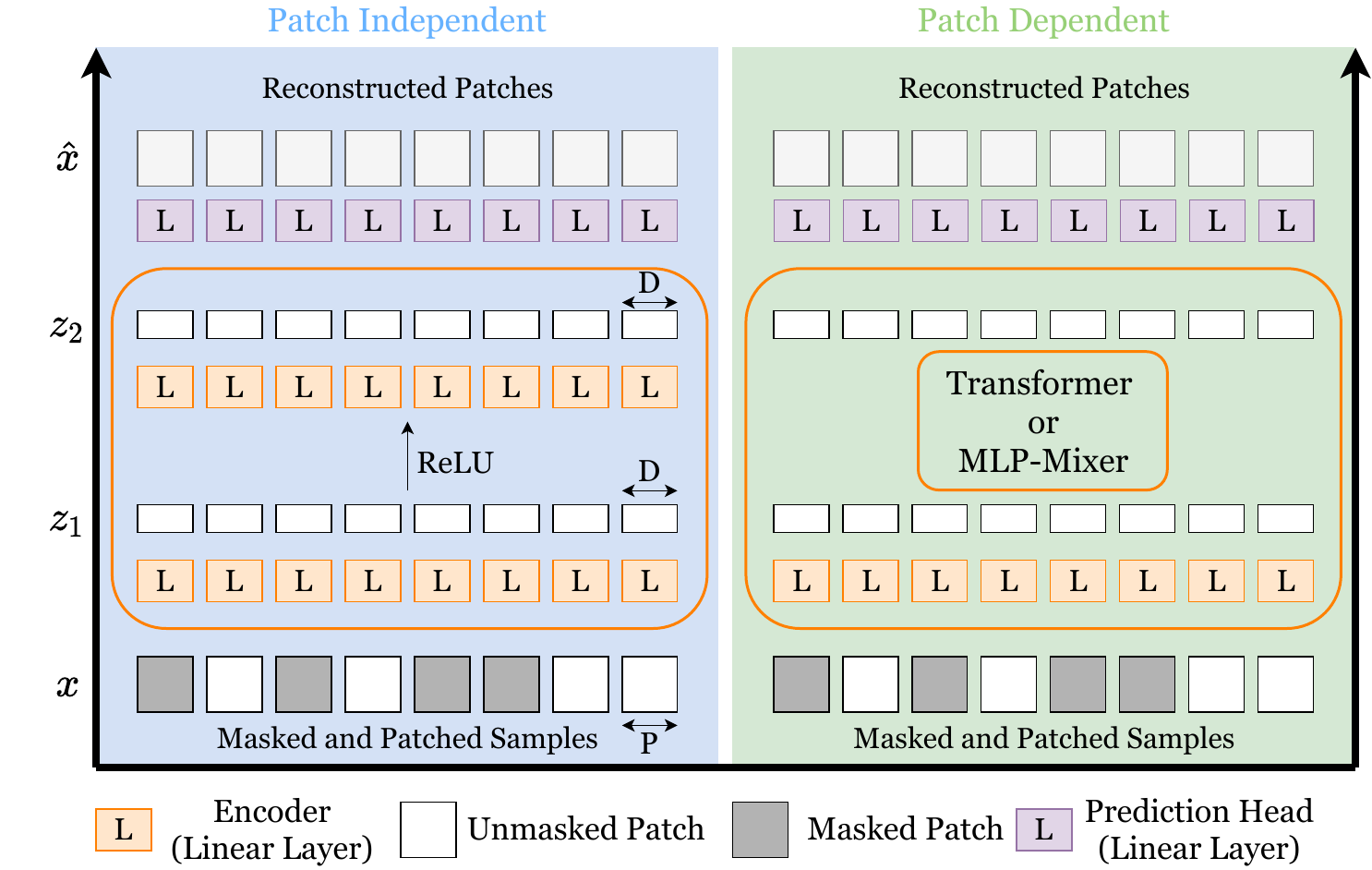}
    \caption{Two model designs for reconstructing masked patches. The left architecture learns each patch independently, while the right architecture allows information sharing among patches.}
    \label{fig:patch_indepence}
\end{figure}

\subsection{Dataset Description}
Our study utilizes a set of publicly available datasets to evaluate performance across various wireless signal recognition tasks (short-range and long-range technologies), \ac{uwb} channel impulse responses, and common modulation schemes. For short-range analysis, two datasets were used. The first \cite{FONTAINE2019101881} contains LTE, Wi-Fi, and DVB-T signals captured in Ghent, Belgium, using an Anritsu MS 2690A signal analyzer at a 10 MHz sampling rate across multiple bands (806 MHz for LTE, 2412/5540 MHz for Wi-Fi, and 482 MHz for DVB-T). The second short-range dataset \cite{b82jsy5723} provides Wi-Fi (IEEE 802.11ax), LTE, and 5G-NR signals captured with a USRP B210 SDR at a 20 MHz sampling rate, with center frequencies of 5.825 GHz, 1.8425 GHz, and 628 MHz, respectively. For long-range technologies, we used the dataset from \cite{9348566}, which focuses on sub-GHz signals, including Sigfox, LoRa, IEEE 802.15.4g, and IEEE 802.11ah, captured using RTL-SDR dongles at a 1 MHz sampling rate. To evaluate \ac{los} and \ac{nlos} detection, we utilized the \ac{uwb} \ac{cir} dataset from \cite{10195942}, which contains approximately 80,000 \ac{uwb} ranging measurements labeled as \ac{los} or \ac{nlos} from industrial, office, and lab environments. Finally, for modulation classification, we used the RADIOML2016.10A dataset \cite{o2016convolutional}, which comprises 11 common analog and digital modulation schemes (e.g., BPSK, QPSK, 16QAM, WB-FM) sampled at approximately 1 MSample/sec and segmented into 128 µs intervals.

\section{Methodology} \label{sec:methodology}

This section details the proposed method. The primary objective is to learn an embedding function \( f_{\theta} \) for \textbf{a patch} \( x_{p}^{(i,c,n,u)}  \) into a representation \( z^{(i,c,n,u)} \), where \( i, c, n, u \) represent the indices for the time, channel (I or Q), patch, and data sample respectively. The input patch \( x_{p}^{(i,c,n,u)} \in \mathbb{R}^{P} \) has dimension \( P \), and the output embedding \( z^{(i,c,n,u)} \in \mathbb{R}^{D} \) has dimension \( D \). The model employs channel independence \cite{10529618}, where a single model's weights are shared across all channels, and \ac{pi} \cite{d13935022a9b4914aa1194be6b069f60}, where weights are also shared across all patches, making the embedding function \( f_{\theta} \) independent of channel and patch indices, as the comparison is depicted in Fig.~\ref{fig:patch_indepence}.

\subsection{Pre-training Strategy}
Instead of the conventional masked modeling approach, which predicts masked patches from unmasked ones (a \ac{pd} task), we use a \ac{pi} task. This task involves autoencoding each patch individually, without reference to other patches. The reconstruction is performed patch-wise using a shared fully connected layer: \( \hat{x}_{p}^{(i,c,n,u)} = Wz^{(i,c,n,u)} \), where \( W \in \mathbb{R}^{P \times D} \).

\subsection{Encoder Architecture}
To align with the \ac{pi} principle, we use a simple \ac{mlp} architecture as the encoder. We use a lightweight \ac{mlp} encoder with shared weights across all channels and patches, ensuring parameter efficiency and patch-level invariance during pre-training. This contrasts with Transformer-based models that are designed to capture dependencies between patches. The proposed \ac{mlp} consists of two layers with a ReLU activation function, as depicted in Fig.~\ref{fig:patch_indepence}.

\subsection{Objective}

\begin{figure}[t]
    \centering
    \includegraphics[width=1\linewidth]{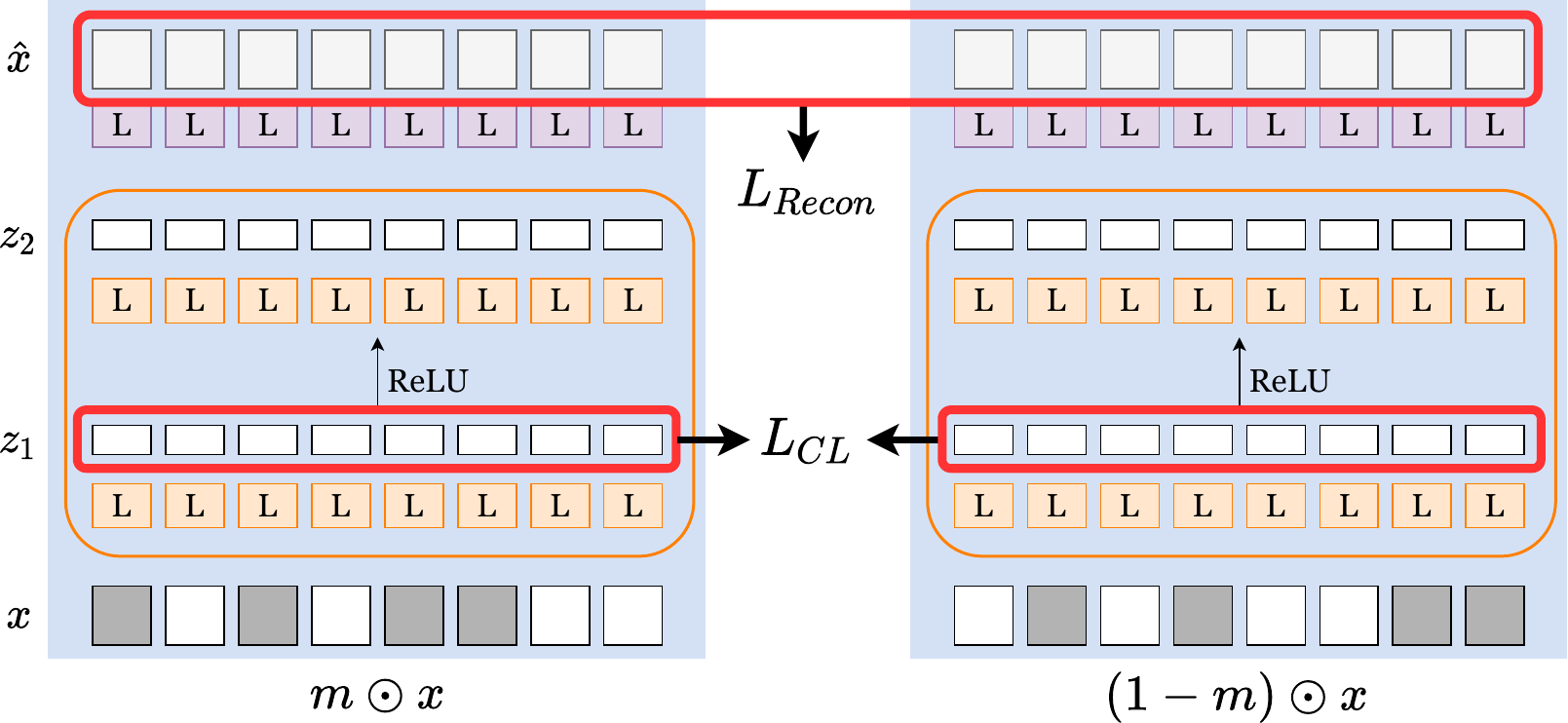}
    \caption{Illustration of contrastive learning with masked views. Two complementary masked inputs are encoded, compared via \(\mathcal{L}_{\text{CL}}\), and decoded with \(\mathcal{L}_{\text{Recon}}\).}
    \label{fig:Loss_model}
\end{figure}

To enhance the representations, we use a complementary \ac{cl} mechanism. This is achieved by creating two augmented views of the input \( x \) using a complementary mask \( m \) array. To generate the two views, we apply a complementary mask with a ratio of 50\%. The two views are \( m \odot x \) and \( (1-m) \odot x \) as depicted in Fig.~\ref{fig:Loss_model}. The \ac{cl} is applied hierarchically \cite{Yue_Wang_Duan_Yang_Huang_Tong_Xu_2022} by max-pooling patch representations along the temporal axis, enabling the model to capture information at different granularities.

The \ac{cl} loss is computed on the representations from the first \ac{mlp} layer, \( z_1 \), and the reconstruction is performed on the representations from the second layer, \( z_2 \), followed by a projection head.

The \textbf{reconstruction loss} is the mean squared error between the original patches and the reconstructed ones:

\begin{equation}
     \mathcal{L}_{\text{Recon}} = \sum_{i=1}^{B}\sum_{c=1}^{C}\sum_{n=1}^{N} \left\| x_{p}^{(i,c,n,u)} - \hat{x}_{p}^{(i,c,n,u)} \right\|_{2}^{2}, 
\end{equation}

\noindent where $B$ is the number of time samples, $C$ is the number of channels, and $N$ is the number of patches per channel.

The \textbf{contrastive loss} uses a softmax formulation over the dot product similarity of patch embeddings. For a pair of views of a patch embedding, \( z_{1}^{(i,c,n,u)} \) and \( z_{1}^{(i,c,n+N,u)} \), the softmax probability is:

\begin{equation}
     p(i,c,(n,n'),u) = \frac{\exp(z_{1}^{(i,c,n,u)} \circ z_{1}^{(i,c,n',u)})}{\sum_{s=1, s \neq n}^{2N} \exp(z_{1}^{(i,c,n,u)} \circ z_{1}^{(i,c,s,u)})}
\end{equation}
The total contrastive loss is the negative log-likelihood over positive pairs:

\begin{equation}
    \mathcal{L}_{\text{CL}} = \frac{1}{2BCN} \sum_{i=1}^{B}\sum_{c=1}^{C}\sum_{n=1}^{2N} -\log p(i,c,(n,n+N),u) 
\end{equation}

Positive pairs refer to the representations of the same patch under two different masking views. Specifically, \( z_1^{(i,c,n,u)} \) and \( z_1^{(i,c,n+N,u)} \) are embedding vectors generated from the same original patch, where different portions of the input have been masked. These complementary masked views share the same semantic information but differ in appearance.

The \textbf{final loss} is the sum of two losses:
\begin{equation}
    \mathcal{L} = \mathcal{L}_{\text{Recon}} + \mathcal{L}_{\text{CL}}
\end{equation}

To handle distribution shifts between training and testing data, instance normalization is applied to each sample before processing.

\subsection{Implementation}
For the self-supervised pre-training, we use a large unlabeled dataset, to learn time series representations. We train the model for 100 epochs. The pre-training phase involves reconstructing individual patches. Key hyperparameters during this phase are the hidden dimension of the MLP encoder of 64, the patch size of 128, the stride of 32, and the input size of 4096. The inputs smaller than 4096 are UWB and modulation. \ac{uwb} \ac{cir} samples are zero-padded on the right side, and modulation samples are upsampled to fit in 4096 meaningful samples. We use NVIDIA A40 GPU with 44 GB RAM \cite{gpulab_idlab}.

For fine-tuning, we adapt the pre-trained model for specific classification downstream tasks of long-range and short-range technology recognition, \ac{uwb} \ac{los}/\ac{nlos}, and modulation recognition. This process involves attaching a classification head—comprising max pooling, dropout, and linear layers—to the pre-trained model. The fine-tuning strategy consists of two stages: first, \ac{lp}, where only the new classification head is trained for 10 epochs, followed by end-to-end fine-tuning, where the entire network is trained. This approach allows the model to effectively transfer the general representations learned during pre-training to the specific nuances of the labeled downstream task. The fine-tuning is conducted for 200 epochs.

The learning rate range test from \cite{7926641} is employed to select optimal learning rates. A higher rate of 0.001 is used during pre-training to quickly learn general features. For fine-tuning, smaller task-specific learning rates are applied: 0.0007 for short-range, 0.0006 for long-range and modulation, and 9.77e-5 for \ac{uwb}, reflecting differences in task complexity. These tailored values help preserve pre-trained representations while adapting to each downstream task.

\section{Results} \label{sec:Results}

\subsection{Performance Analysis}

This section presents the performance evaluation of the proposed \ac{pi} \ac{mlp} model, with a focus on its fine-tuning capabilities across various downstream tasks and its robustness when encountering previously unseen data classes. The results are benchmarked against SpectrumFM \cite{zhou2025spectrumfmfoundationmodelintelligent}, IQFM \cite{mashaal2025iqfmwirelessfoundationalmodel}, and a \ac{pd} Transformer model \cite{cheraghinia2025foundationmodelwirelesstechnology}.

\begin{table}[!htbp]
\centering
\begin{tabular}{m{2.3cm} m{0.7cm} m{1.1cm} m{1.1cm} m{1.2cm}}
\hline
\textbf{Models} & \textbf{Long-range} & \textbf{Short-range} & \textbf{UWB} & \textbf{Modulation} \\
\hline
 Fontaine et al.~\cite{9348566} & \textbf{100\%} & - & - & - \\
Fontaine et al.~\cite{10195942} & - & - & $\leq$ 81.5\% & - \\
 SpectrumFM \cite{zhou2025spectrumfmfoundationmodelintelligent} & - & $\approx$ 82.5\% & - & \textbf{$\approx$ 92\%} \\
IQFM \cite{mashaal2025iqfmwirelessfoundationalmodel} & - & - & - & \textbf{$\approx$ 99\%} \\
PD Transformer \cite{cheraghinia2025foundationmodelwirelesstechnology} & 99.53\% & 86.03\% & 97.32\% & 52.1\% \\
\hline
PI MLP & \textbf{100\%} & \textbf{97.25\%} & \textbf{97.37\%} & 63.25\% \\

\hline
\end{tabular}
\caption{Fine-tuning results for downstream tasks and comparison to other approaches. Although our model has fewer parameters, it outperforms prior models.}
\label{tab:PD_vs_PI}
\end{table}

Table~\ref{tab:PD_vs_PI} presents the fine-tuning performance of our proposed \ac{pi} \ac{mlp} model across four distinct downstream tasks, alongside results from other relevant models in the literature. The \ac{pi} \ac{mlp} consistently achieves superior or competitive accuracy in all tasks compared to the \ac{pd} Transformer. Notably, the \ac{pi} \ac{mlp} attains perfect accuracy (100\%) in long-range technology recognition, surpassing the \ac{pd} Transformer’s 99.53\% and matching the task-specific model from \cite{9348566}, which focuses solely on this task.

For short-range recognition, the \ac{pi} \ac{mlp} demonstrates a significant improvement with 97.25\% accuracy, outperforming both the \ac{pd} Transformer (86.03\%) and the approximate 82.5\% reported by SpectrumFM \cite{zhou2025spectrumfmfoundationmodelintelligent}. For UWB LOS/NLOS detection, the \ac{pi} \ac{mlp} achieves 97.37\% accuracy, slightly exceeding the \ac{pd} Transformer’s 97.32\% and outperforming the non-foundation model in \cite{10195942}, which reports up to 81.5\%. 

Regarding the more challenging modulation recognition task —involving 11 classes — the \ac{pi} \ac{mlp} attains 63.25\% accuracy, improving upon the \ac{pd} Transformer’s 52.1\%. However, it still falls short of the higher accuracies reported by foundational models such as SpectrumFM (\(\approx\) 92\%) and IQFM (\(\approx\) 99\%) \cite{zhou2025spectrumfmfoundationmodelintelligent, mashaal2025iqfmwirelessfoundationalmodel}. A key limitation observed is the reduced performance of the \ac{pi} \ac{mlp} on tasks involving shorter data sequences, like modulation recognition. This suggests that current model designs struggle to fully capture the complexities of such variable-length sequences in wireless data, which are common in wireless data because signals are often recorded using different sampling rates. Similarly, larger foundation models such as SpectrumFM also face performance trade-offs, excelling in some tasks but underperforming in others, such as technology recognition, despite their scale. These observations highlight that merely increasing model size is not sufficient to address the challenges posed by diverse wireless datasets, emphasizing the need for more nuanced architectural or training strategies in future work, especially for data input sources with variable lengths due to different sampling rates, different start/stop differences, etc.

\begin{table}[H]

    \centering
    \begin{tabular}{lcccc}
        \toprule
        \textbf{Models} & \multicolumn{1}{c}{\textbf{LTE Excluded}} & \multicolumn{1}{c}{\textbf{Zigbee Excluded}} \\
        \cmidrule(lr){2-3} \cmidrule(lr){4-5}
        & Short-range  & Long-range \\
        \midrule
        PD Transformer \cite{cheraghinia2025foundationmodelwirelesstechnology}  & 90.12\% & 99.99\% \\
	PI MLP  & \textbf{97.55\%} & \textbf{100\%} \\
        \bottomrule
    \end{tabular}
        \caption{Accuracy of downstream tasks for classes that were unseen during pre-training. }
    \label{tab:finetuning_unseen}
\end{table}

To evaluate the model's generalization capacity, we conducted experiments where specific wireless technologies were excluded from the pre-training dataset and introduced only during the fine-tuning phase. Table~\ref{tab:finetuning_unseen} presents these findings. When LTE was excluded, the \ac{pi} \ac{mlp} maintained a high accuracy of 97.55\% on the short-range task, outperforming the \ac{pd} Transformer, whose focus on learning dependencies between signal patches; these relational features can be less discriminative and fail to generalize to the unseen class that has different sampling rates. For example, the LTE class contains samples from two datasets with different sampling rates, resulting in a drop in accuracy to 90.12\%. Similarly, when Zigbee was the unseen class, the \ac{pi} \ac{mlp} achieved 100\% accuracy for the long-range task, while the \ac{pd} Transformer reached 99.99\%, considering that Zigbee has a single sampling rate. These results underscore the \ac{pi} \ac{mlp}'s ability to generalize and adapt to new, previously unobserved classes, a critical feature for scalable and real-world deployable foundation models, by learning more robust and intrinsic patch features. 

\begin{figure}[!htbp]
    \centering
    \includegraphics[width=0.97\linewidth]{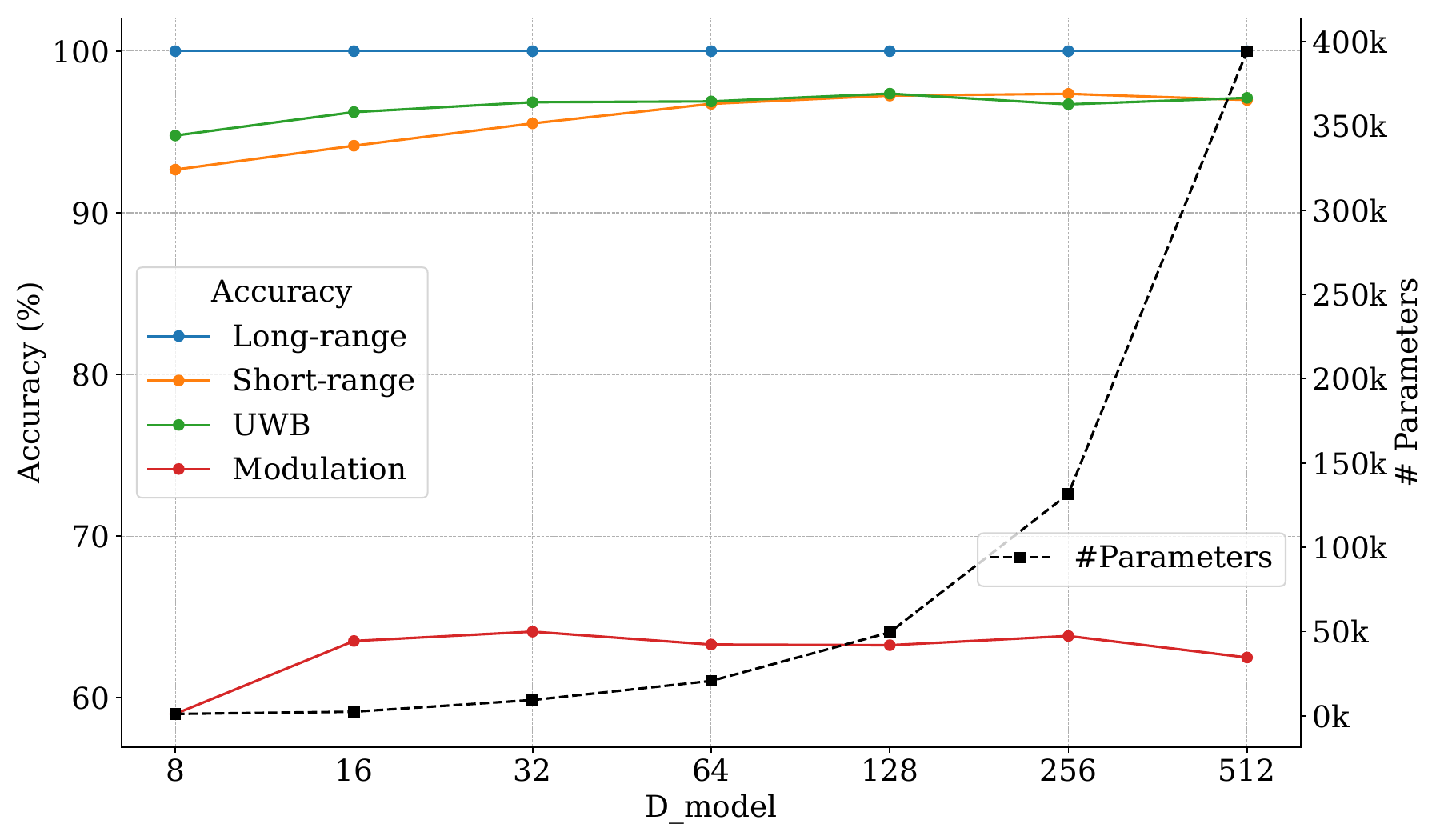}
    \caption{Impact of the encoder hidden dimension ($D_{\text{model}}$) on the number of trainable parameters and fine-tuning accuracy for four downstream tasks.}
    \label{fig:d_mode_comparison}
\end{figure}

Fig.~\ref{fig:d_mode_comparison} illustrates the impact of varying the encoder hidden dimension, $D_{\text{model}}$, on both model complexity (number of parameters) and fine-tuning accuracy across the four downstream tasks. For this analysis, the patch size and stride were held constant at 128 and 32, respectively. The black dashed line clearly shows that the number of trainable parameters grows substantially as the encoder dimension increases, rising from under 1K to nearly 400K. Despite the substantial increase in model complexity, the accuracy of downstream tasks remains largely unchanged when the dimension exceeds 64. Performance on long-range recognition is stable at 100\%, while short-range and UWB detection accuracies also remain consistently high, fluctuating minimally around 97\%. Similarly, the accuracy for modulation recognition, the most challenging task, hovers around 63-64\% with no meaningful improvement resulting from a larger model dimension.This analysis shows that increasing model dimension does not cause significant performance gains. This is due to the choice of a lightweight MLP encoder. To achieve better performance, other lightweight encoders will be studied in the future.

\begin{table}[!htbp]
\centering
\begin{tabular}{lcccc}
\hline
\textbf{Method} & \textbf{Long-range} & \textbf{Short-range} & \textbf{UWB} & \textbf{Modulation} \\
\hline
LP & 98.68\% & 59.7\% & 84.85\% & 48.41\% \\
FN & \textbf{100\%} & 97.1\% & 96.95\% & 62.6\% \\
LP-FN & \textbf{100\%} & \textbf{97.25\%} & \textbf{97.37\%} & \textbf{63.25\%} \\

\hline
\end{tabular}
\caption{Impact of different fine-tuning strategies on downstream task accuracy}
\label{tab:LP_FN_comp}
\end{table}

Table~\ref{tab:LP_FN_comp} shows the comparison of three fine-tuning strategies — LP, FN, and LP-FN — across four downstream tasks. The results demonstrate that fine-tuning improves accuracy over linear probing alone, especially in short-range, UWB, and modulation tasks. Moreover, combining LP and FN (LP-FN) produces the best performance, demonstrating the robustness benefits of integrating both approaches in fine-tuning \cite{tomihari2024understanding}.

\subsection{Complexity Analysis}

This section provides a complexity analysis of the proposed \ac{pi} \ac{mlp} model in comparison to the \ac{pd} Transformer and other relevant foundation models, with key metrics such as pre-training time, number of parameters, and inference time as mentioned in Table~\ref{tab:model_comparison}. The critical advantage of our approach is the exceptionally fewer number of parameters, which directly translates to computational efficiency.

As detailed in Table~\ref{tab:model_comparison}, the \ac{pi} \ac{mlp} comprises only $\approx$21K parameters. This renders it significantly more compact than the \ac{pd} Transformer ($\approx$700K), IQFM ($\approx$341K), and especially SpectrumFM ($\approx$30.7M). The training time further underscores this efficiency, with the \ac{pi} \ac{mlp} requiring 2 minutes per epoch for pre-training, compared to 5 minutes for the \ac{pd} Transformer. \textbf{Our model's lightweight architecture makes it the most suitable for deployment on resource-constrained edge devices}, enabling advanced downstream tasks without demanding extensive memory or processing power.

To further validate the inference, we measured the average inference time on an NVIDIA Jetson Xavier NX device. After a 10-step warm-up, we evaluated 5000 samples. The reported inference times reflect this setup. These measurements confirm that the \ac{pi} \ac{mlp} achieves significantly faster inference, making it well-suited for real-time applications on edge hardware.

\begin{table}[!htbp]
\centering
\begin{tabular}{l  m{2 cm}  m{1.6 cm}  m{1.3 cm}}
\hline
\textbf{Models} & \textbf{Pre-training time (per epoch)} & \textbf{\# parameters}& \textbf{Inference time}\\
\hline
SpectrumFM \cite{zhou2025spectrumfmfoundationmodelintelligent}    & Not mentioned  & $\sim$30.7M & 62.73 ms\\
IQFM \cite{mashaal2025iqfmwirelessfoundationalmodel}    & Not mentioned  & $\sim$341K & 1.22 ms\\
PD Transformer \cite{cheraghinia2025foundationmodelwirelesstechnology} & 5 min          & $\sim$700K & 0.847 ms\\
\hline
PI MLP     & \textbf{2 min} & $\sim$\textbf{21K} & \textbf{0.33 ms} \\
\hline

\end{tabular}
\caption{Model complexity comparison by pre-training time, trainable parameters, and inference time.}
\label{tab:model_comparison}
\end{table}

\section{Conclusion \& Future Works} \label{sec:Conc}
This paper introduced a lightweight foundation model based on a \ac{pi} MLP architecture for wireless time-series tasks on edge devices. Our results demonstrate that, despite its compact size (21K parameters), the model achieves competitive performance across four diverse downstream tasks, including generalization to previously unseen wireless technologies. Notably, the model achieves an average inference time of just 0.33 milliseconds on an NVIDIA Jetson Xavier NX, allowing real-time signal analysis. These findings challenge the reliance on large Transformer-based models and highlight the potential of simpler architectures for scalable, low-power deployment in real-world wireless systems.

Future work will focus on improving performance for tasks involving short-length sequences, such as modulation recognition, by supporting dynamic patch sizes. We also aim to extend the model’s applicability to broader wireless modalities and explore hybrid architectures, balancing explainability with efficiency and representational capacity to enhance trust.

\bibliographystyle{unsrt}
\bibliography{References}

\end{document}